\documentclass{PoS}

\usepackage{cleveref}
\usepackage{url}
\usepackage{gensymb}
\usepackage{graphicx}
\usepackage{subcaption}
\usepackage{multirow}
\usepackage{wrapfig}
\usepackage{multicol}

\newcommand{\GeV}{\mathrm{GeV}}
\newcommand{\cm}{\mathrm{cm}}
\newcommand{\s}{\mathrm{s}}

\newcommand{\point}{\textit{Point}}
\newcommand{\ext}{\textit{Extended}}
\newcommand{\on}{\texttt{ON}}

\newcommand{\veritas}{{VERITAS}}

\newcommand{\hess}{{H.E.S.S.}}
\newcommand{\fermi}{\textit{Fermi}}
\newcommand{\hawc}{{HAWC}}

\newcommand{\LAT}{\textit{Fermi}-LAT}
\newcommand{\milagro}{{Milagro}}

\newcommand{\xmm}{\textit{XMM-Newton}}

\newcommand{\suzaku}{\textit{Suzaku}}

\newcommand{\cisne}{VER J2019\allowbreak+368}
\newcommand{\TeVJ}{VER J2031\allowbreak+415}
\newcommand{\GCyg}{VER J2019\allowbreak+407}
\newcommand{\CTB}{VER J2016\allowbreak+371}

\newcommand{\MGROcisne}{MGRO J2019\allowbreak+37}

\newcommand{\cisneA}{VER J2018\allowbreak+367}
\newcommand{\cisneB}{VER J2020\allowbreak+368}

\newcommand{\fermipy}{\emph{fermipy}}

\newcommand{\GR}{$\gamma$-ray}

\newcommand{\No}{$N_0$}
\newcommand{\Eo}{$E_0$}

\newcommand{\ra}{$\alpha_{J2000}$}
\newcommand{\dec}{$\delta_{J2000}$}

\newcommand{\hms}[3]{$#1^\mathrm{h}\allowbreak#2^\mathrm{m}\allowbreak#3^\mathrm{s}$}

\newcommand{\dms}[3]{$#1\degree\allowbreak#2'\allowbreak#3''$}

\newcommand{\dstat}[2]{$#1\degree\allowbreak\pm#2^\circ_{stat}$}
\newcommand{\dee}[3]{$#1\degree\allowbreak\pm#2^\circ_{stat}\allowbreak\pm#3^\circ_{sys}$}

\newcommand{\nstat}[2]{$#1\allowbreak\pm#2_{stat}$}
\newcommand{\nee}[3]{$#1\allowbreak\pm#2_{stat}\allowbreak\pm#3_{sys}$}

\newcommand{\fstat}[3]{$(#1 \pm #2_{stat})\allowbreak \times 10^{#3} \, \allowbreak \GeV^{-1}\cm^{-2}\s^{-1}$}
\newcommand{\fee}[4]{$(#1\allowbreak\pm#2_{stat}\allowbreak\pm#3_{sys} )\allowbreak \times  10^{#4}\allowbreak \, \mathrm{GeV}^{-1}\allowbreak  \mathrm{cm}^{-2}\allowbreak  \mathrm{s}^{-1}$}

\title{VERITAS observations of the Cygnus Region}

\ShortTitle{VERITAS observations of the Cygnus Region}

\author{\speaker{Ralph Bird}, for the \veritas\ Collaboration\thanks{http://veritas.sao.arizona.edu}\\
        University of California, Los Angeles\\
        E-mail: \email{ralphbird@astro.ucla.edu}}

\abstract{The Cygnus region of the galaxy is one of the richest regions of gas and star formation and is the brightest region of diffuse GeV emission in the northern sky. VERITAS has conducted deep observations (approximately 300 hours) in the direction of Cygnus region, reaching an average sensitivity of a few percent of the Crab nebula flux. We present the results of these observations and an analysis of over seven years of \LAT\  data above 1~GeV. In addition to a search for new sources in the region, we present updated spectra and morphologies of the known TeV \GR\ sources and a study of their relationship with the GeV emission from the region. These results are discussed in their multiwavelength context including the recently published HAWC observatory \GR\ catalog. A comparison is also made to the \hess\ galactic plane survey.}

\FullConference{35th International Cosmic Ray Conference\\
		 10-20 July, 2017\\
		 Bexco, Busan, Korea}

\begin{document}
\include{abbrev}

\section{Introduction}
The Cygnus region is the brightest region of diffuse high-energy (HE, 0.1 GeV $<$ E $<$ 100 GeV) \GR s in the northern sky. 
Seen as a small-scale version of a whole galaxy, the Cygnus region harbors a wealth of objects and is the largest known star-forming region outside the Galactic center. 
It has already been observed by a wealth of instruments at different wavelengths which have highlighted the variety of objects and processes within the region, and firmly established it as a key region for understanding our galaxy.

We present a survey over a 15\degree\ by 5\degree\ portion of the Cygnus region centered on Galactic longitude ($l$) 74.5\degree\ and Galactic latitude ($b$) 1.5\degree\ conducted by \veritas\ between 2007 April and 2008 December with a total observing time of 135 h (120 h live time). 
We also include targeted and follow-up observations of 174~h (151~h live time) made by \veritas\ between 2008 November and 2012 June, for a total observing time of about 309 h (271 h live time).


The Cygnus region has been observed by several very high energy (VHE, E $>$ 100~GeV) \GR\ instruments who have identified seven sources, four of them have previously been detected by \veritas.
TeV J2032+4130 is an unidentified VHE emitter which lies within the extended Cygnus cocoon \cite{VERTeV}.
\GCyg\ is also located within the Cygnus cocoon and is associated with the Gamma-Cygni SNR (G78.2+2.1) \cite{VERGCyg}. 
The large, bright \milagro\ source \MGROcisne\ which has been resolved into two sources after observations by \veritas: \CTB\ which is associated with the PWN CTB 87, and \cisne\ which is a spatially extended source whose origin has yet to be identified \cite{VERCisne}.
\hawc\ has recently published the 2HWC catalog \cite{2HWC}, in which they identified three new sources in the survey region: 2HWC~J1953+294, which lies at the edge of the survey region; 2HWC~J2006+341; and 2HWC~J2024+417*.

\section{\LAT\ Observations and Analysis}
The Large Area Telescope (LAT) \cite{LAT}, is the primary instrument on the \textit{Fermi Gamma-ray Space Telescope}.  
It is a pair conversion \GR\ detector that is sensitive to \GR s with energies from 20~MeV to greater than 500~GeV. 
We have undertaken an analysis using over seven years (2008 August - 2016 January) of \LAT\ Pass 8 data using \LAT\ science tools v10r0p5 and the \fermipy\ tools  v0.13.5. 
In order to reduce the contribution of the galactic diffuse emission and for improved angular resolution, ``SOURCE'' class photons were selected in the energy range from 1 to 500~GeV and a region of 30\degree\ radius centered at ($l$, $b$) = (74.5\degree, 1.5\degree). 
The region of interest is taken to be 65.5\degree\ $<$ $l$ $<$ 83.5\degree, -2.5\degree\ $<$ $b$ $<$ 5.5\degree\ to match the \veritas\ data. 
A base model derived from the 3FGL was used with the \fermipy\ tool \emph{find\_sources} used to identify new sources. 
All sources have a minimum test statistic (TS) of 25 and spectral models were updated if appropriate.

By far the brightest sources in the region are pulsars, but they all demonstrate an exponential cutoff spectrum with an extrapolated flux greater than 1~TeV of approximately zero. 
To search for potential PWN emission, a ``gated'' analysis was conducted, where a time cut was applied to the pulsar phase to remove the \on-pulse and bridge emission.
In addition, we performed an \on-pulse analysis on the data range available to fit an \on-pulse source. 
The spectral parameters of the pulsar in the model used to analyze the full dataset were then fixed to the \on-pulse values and the dataset refit.  
Nebula emission should then be apparent as a positive residual. 
A check was also conducted on the light curves of the pulsars to ensure no changes in the flux occurred during the period of these observations.
Provided the new source has a TS of at least 25 after being refit with the pulsar parameters free, it was kept in the model.
This produced the base model from which all analyses were conducted.

\section{\veritas\ Observations and Analysis}
The Very Energetic Radiation Imaging Telescope Array System (\veritas) is an array of four IACTs, located at the Fred Lawrence Whipple Observatory in southern Arizona (31\degree 40'~N, 110\degree 57'~W, 1.3~km a.s.l.) \cite{VERITAS}. 
Full array operations began in 2007 and in the summer of 2009 the first telescope was relocated to increase the sensitivity of the array \cite{VERUp}.
Following a trigger upgrade in fall 2011, in summer 2012, the cameras in each telescope were replaced with new, high quantum efficiency PMTs which has resulted in a decrease of the array energy threshold to about 85~GeV \cite{VERPerf}. 
All the data presented in this work were taken prior to the 2012 upgrade.

The results presented here were generated using a standard \veritas\ event reconstruction package \cite{LSI} and cross checked with another package, using two different integration radii, one targeted at point sources ($\theta_{int}$ = 0.1\degree, \point) and one at extended sources,  ($\theta_{int}$ = 0.23\degree, \ext).
A high signal threshold ($\sim$130 photoelectrons) and a minimum of three telescope images was used to reduce the  PSF and better handle background artifacts due to bright stars.
Source morphologies were determined by fitting a sky map of the uncorrelated excess events with a delta function and two-dimensional Gaussian distributions convolved with the \veritas\ PSF. 

\section{Results}
\subsection{Whole Region}
\begin{figure*}[tb]
\includegraphics[width=1\textwidth]{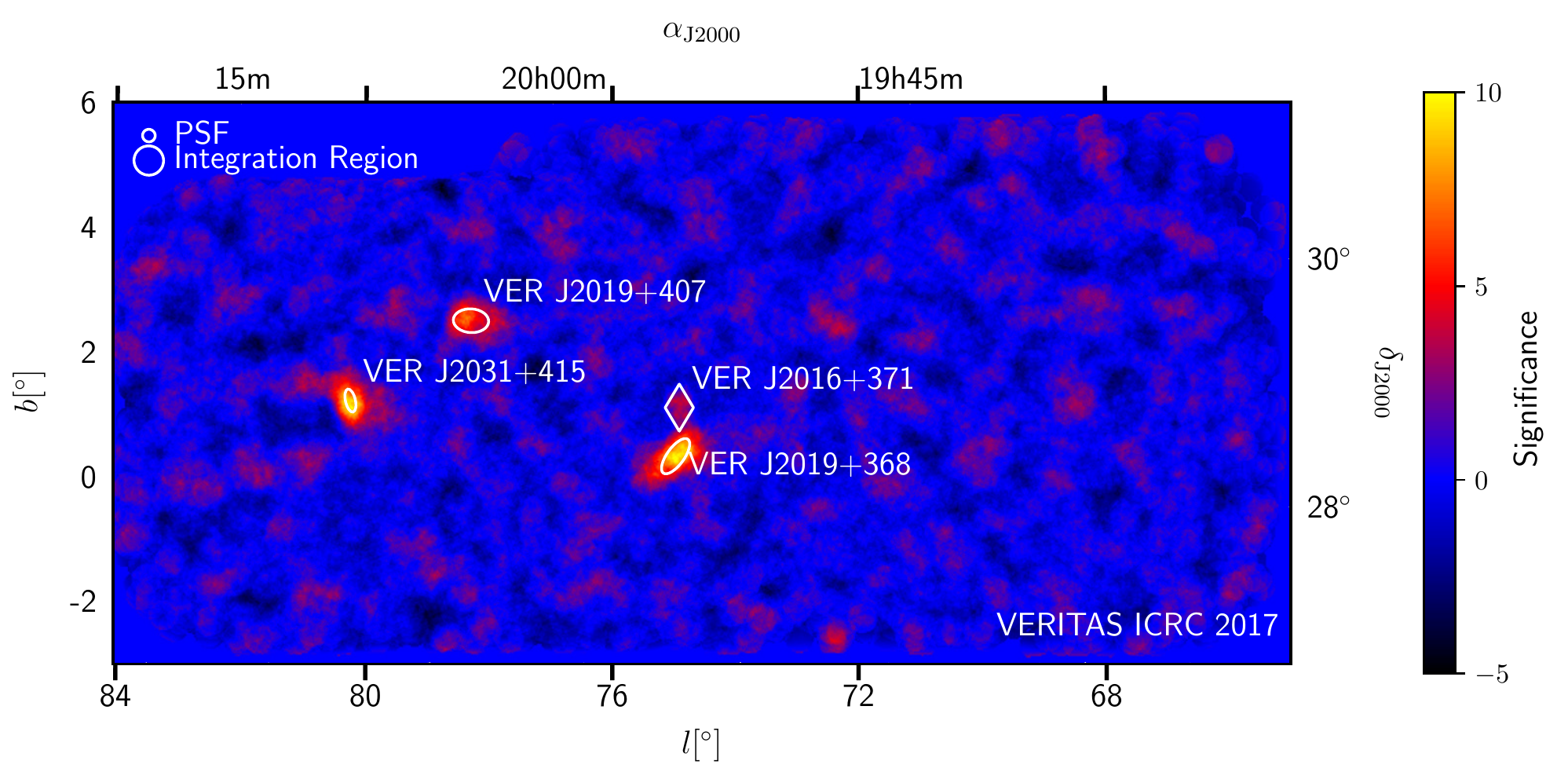}
\caption{Significance map of the entire region using the \ext\ integration region.  
Overlaid are the 1$\sigma$ ellipses for extended sources and the location of \CTB\ (diamond). }
\label{fig:VERSkymaps}
\end{figure*}

Examination of the significance sky map (\cref{fig:VERSkymaps}) produced with the \ext\ integration radius shows the four known \veritas\ sources.
No other new VHE sources were detected.
In addition to the detected VHE sources, upper limits were calculated for 71 locations, including all of the detected \LAT\ sources and other potential VHE emitters in the region.  
The mean significance of the upper limit positions was 0.33$\sigma$ and 0.18$\sigma$ for the \point\ and \ext\ integration radii, respectively.

Twenty-seven 3FGL catalog sources were identified within the region of interest, overlapping with eight 1FHL sources, and four 2FHL sources.
In addition, 25 new point sources were identified.
Two 3FHL catalog sources were not detected in this analysis, 3FHL J1950.5+3457 and 3FHL J2026.7+3449.
Both of these sources are close to the 3FHL detection threshold (5.5$\sigma$ and 4.2$\sigma$ respectively) and hard spectrum (spectral indices of 1.8 and 1.9). 
Of these new \LAT\ sources, notable detections include Cygnus X1/X3, 2HWC J2006+341, PSR J2006+3412 and G73.9+0.9.

\subsection{TeV J2032+4130 / \TeVJ}
\TeVJ\ was observed by \veritas\ at 10.1$\sigma$ using the \ext\ integration radius.
It was found to be asymmetric with a centroid at ($l$, $b$) = \dee{80.25}{0.01}{0.01}, \dee{1.20}{0.01}{0.01} ((\ra, \dec) = \hms{20}{31}{33}, \dms{41}{34}{48})), a semi-major axis of \dee{0.19}{0.02}{0.01}, a semi-minor axis of \dee{0.08}{0.01}{0.03} and oriented to the North-East (of galactic north) at an angle of \dee{13}{4}{1}. 

\begin{figure}[tb]
\centering
\begin{minipage}[t]{0.48\textwidth}
  \centering
\includegraphics[width=\textwidth]{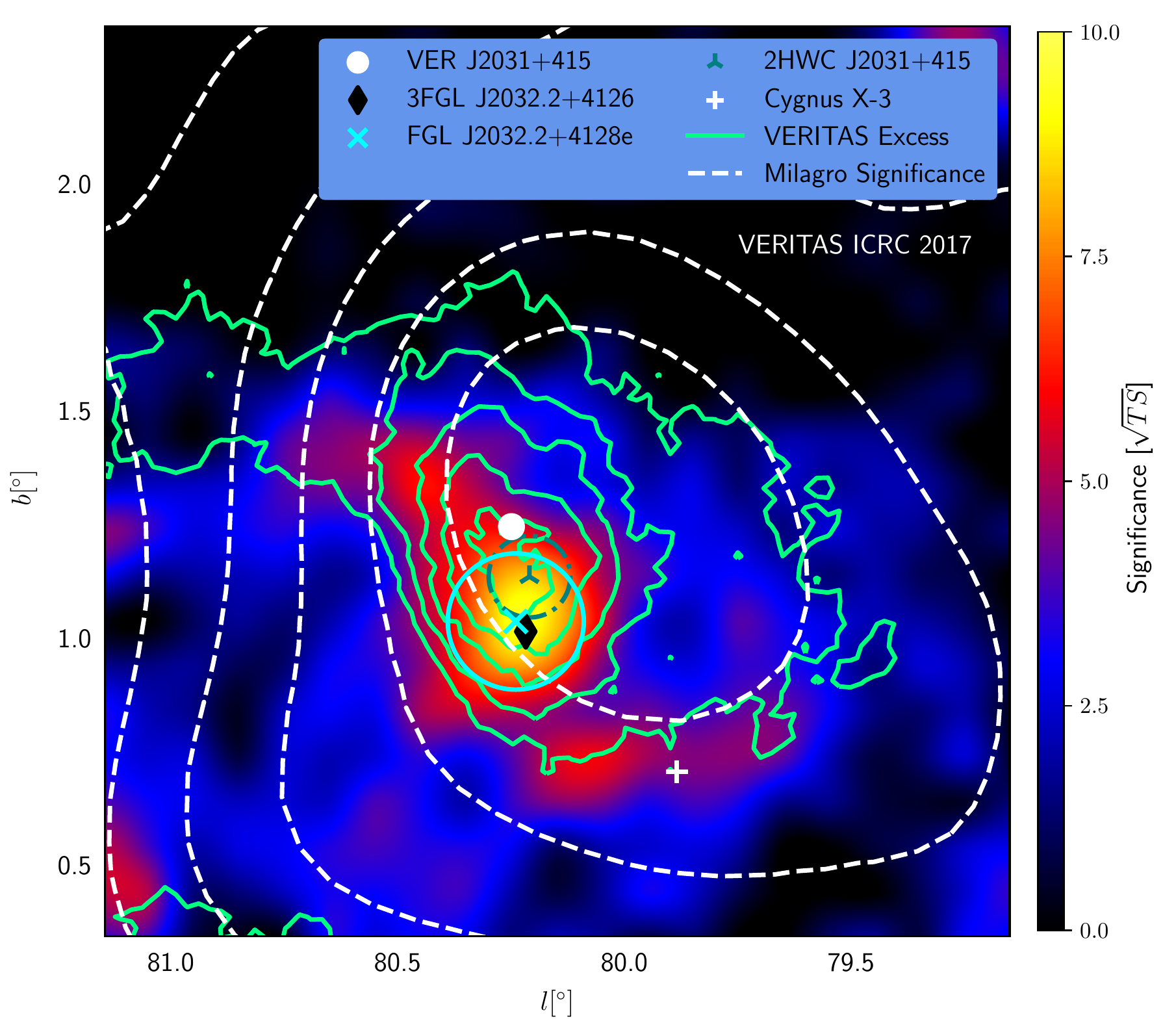}
\caption{Residual map for the \LAT\ observations with PSR J2032+4127 fixed to the \on-pulse parameters. 
\label{fig:TeV2032FermiResid_VERConts}}
\end{minipage}%
\hfill
\begin{minipage}[t]{.48\textwidth}
  \centering
\includegraphics[width=\textwidth]{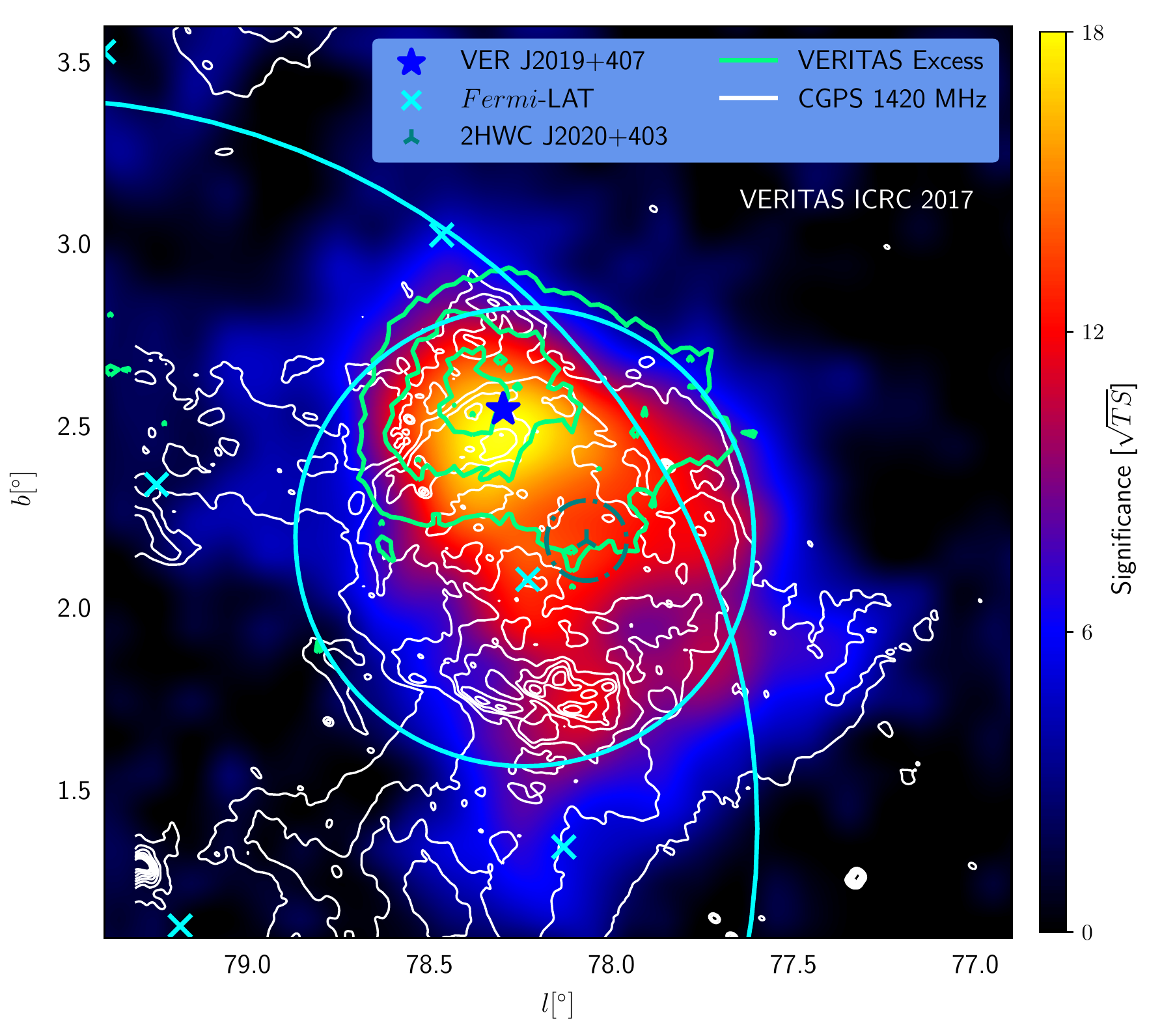}
\caption{\LAT\ $\sqrt{\mathrm{TS}}$ map of 3FGL J2021.0+4031e (SNR G78.2+2.1). 
\label{fig:FermiGammaCyg}}
\end{minipage}
\end{figure}

A gated \LAT\ analysis on 3FGL J2032.2+4126 showed an extended residual (\cref{fig:TeV2032FermiResid_VERConts}), with centroid ($l$, $b$) = (\dstat{80.24}{0.03}, \dstat{1.04}{0.03}) ((\ra, \dec) = (\hms{20}{32}{13}, \dms{41}{28}{39}).
This residual was fit by an symmetric Gaussian source of 68\% containment radius $0.15^{\circ} {}^{+0.02^\circ}_{-0.03^\circ}$ centered on this location with a TS of 321.1 and a TS of extension of 28.6.
We name the source FGL~J2032.2+4128e.

\begin{figure}[tb]
\centering
\begin{subfigure}[t]{.48\textwidth}
  \centering
\includegraphics[width=\textwidth]{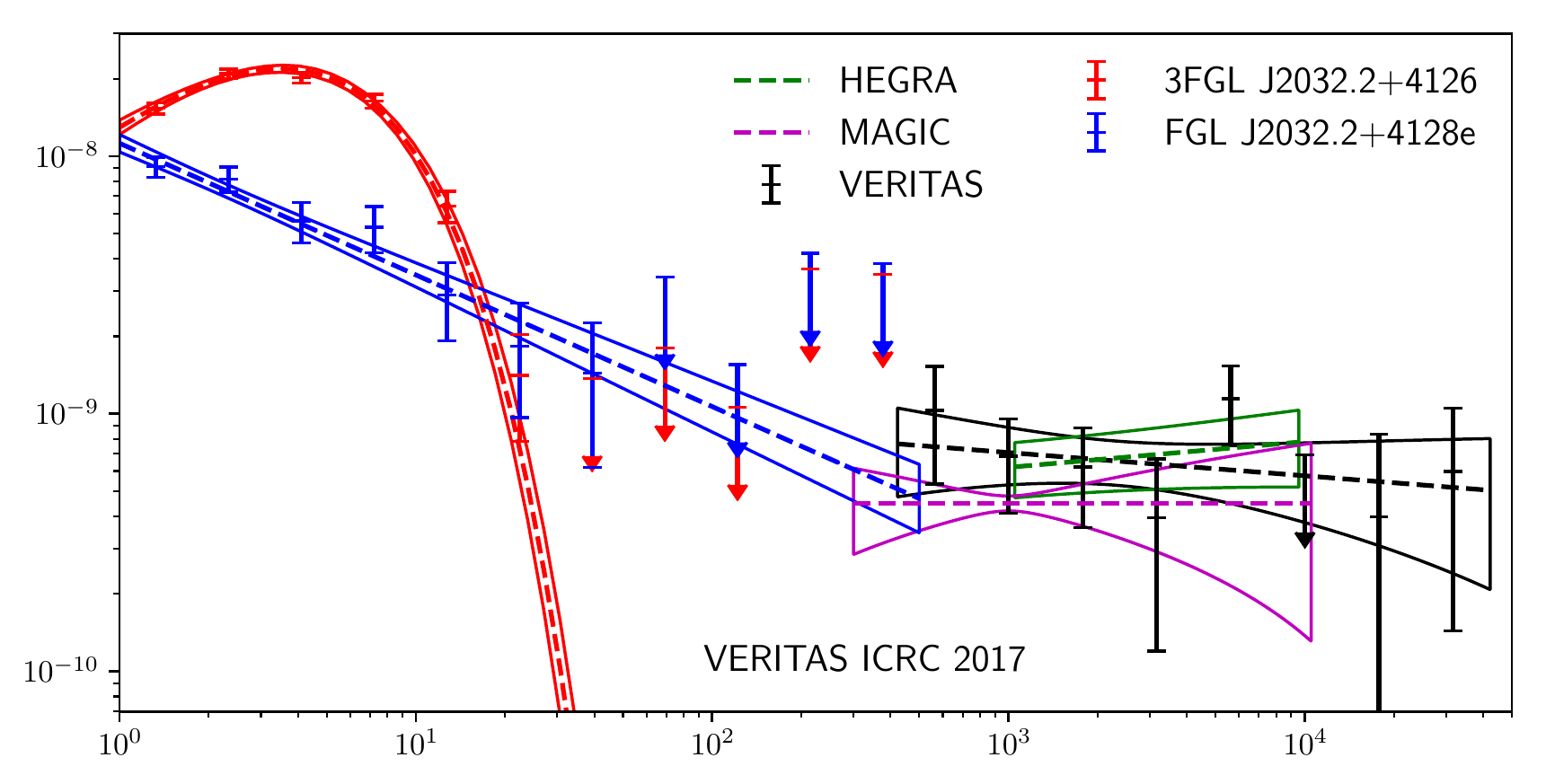}
\caption{TeV J2032+4130 region.  
\label{fig:TeV2032Spectrum}}
\end{subfigure}%
\hfill
\begin{subfigure}[t]{.48\textwidth}
  \centering
\includegraphics[width=\textwidth]{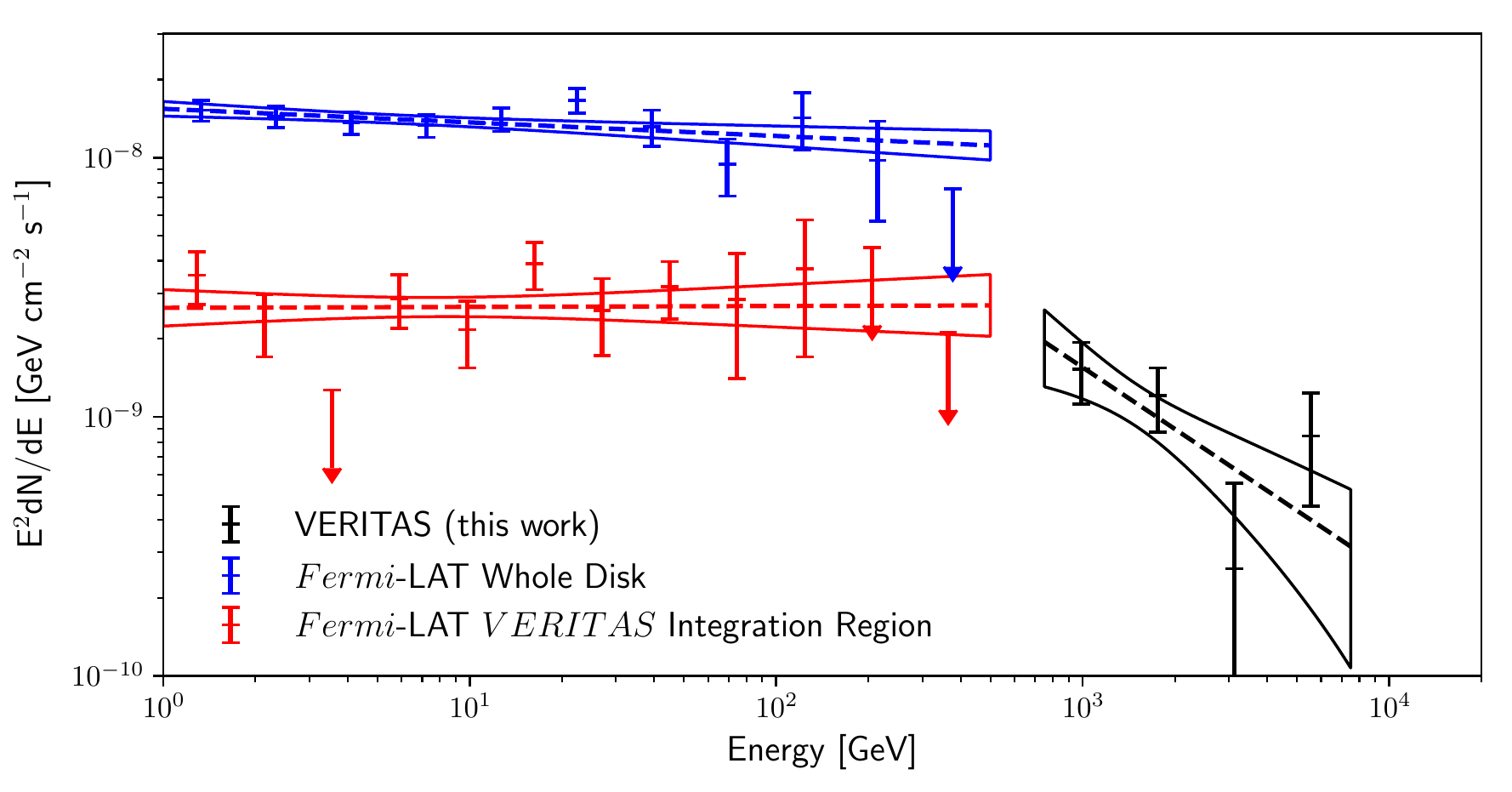}
\caption{Gamma Cygni region.  
\label{fig:GammaCygniSpectrum}}
\end{subfigure}
\begin{subfigure}[t]{.48\textwidth}
  \centering
\includegraphics[width=\textwidth]{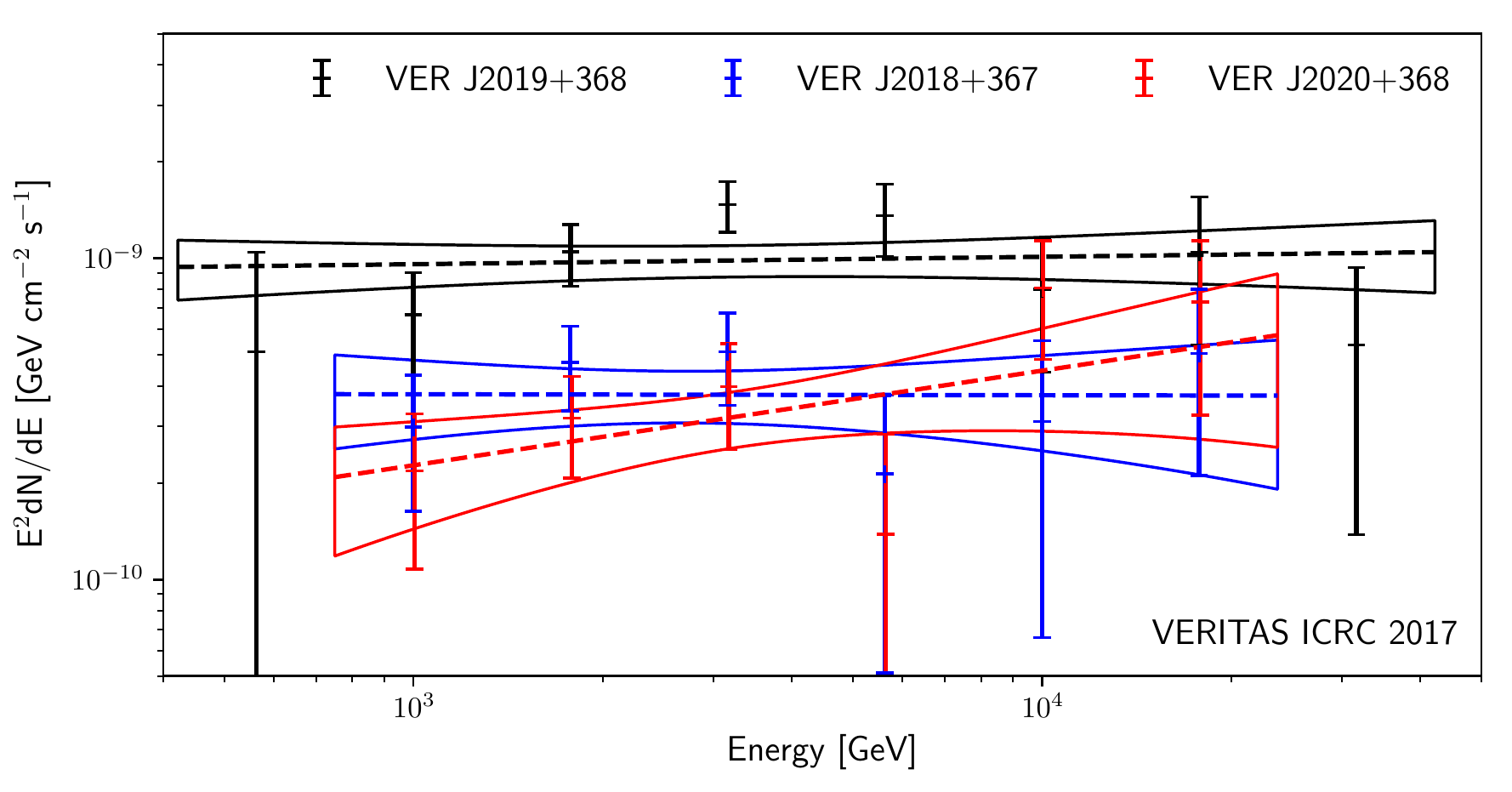}
\caption{\MGROcisne\ region. 
\label{fig:CisneSpectrum}}
\end{subfigure}%
\hfill
\begin{subfigure}[t]{.48\textwidth}
\includegraphics[width=\textwidth]{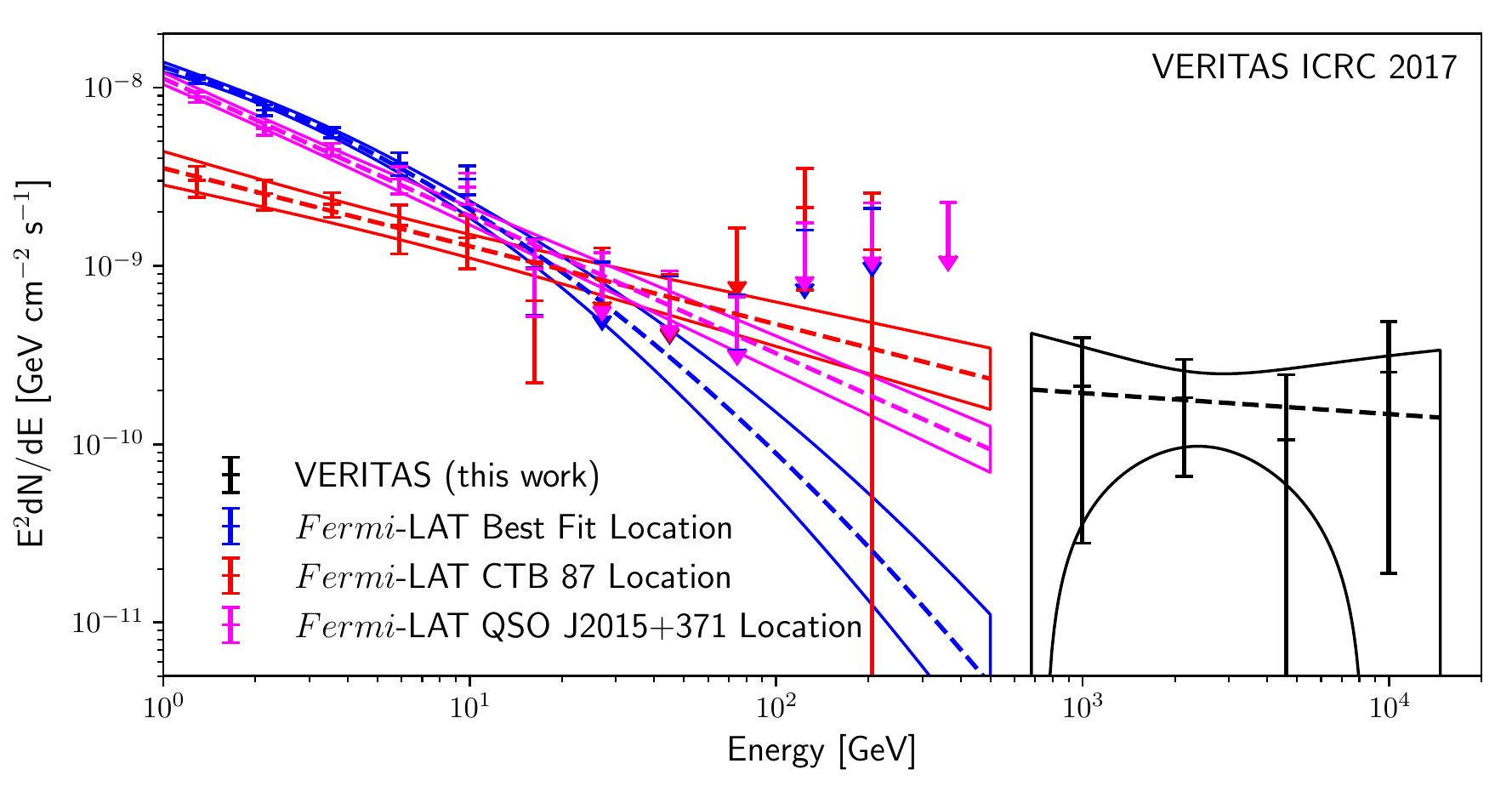}
\caption{CTB 87 region.  
\label{fig:CTB87Spectrum}}
\end{subfigure}
\caption{\LAT\ and \veritas\ spectra.  
\label{fig:Spectrum}}
\end{figure}

The \veritas\ spectrum is well described by a power law with an index of \nee{2.09}{0.19}{0.20} and a normalization of \fee{1.24}{0.24}{0.23}{-16} at 2300 GeV (\cref{fig:TeV2032Spectrum}). 
The spectrum of FGL~J2032.2+4128e can be described by a power law with index \nstat{2.52}{0.07} and a normalization of \fstat{1.44}{0.09}{-8} at 2.27 GeV.
Comparing the spectra of FGL~J2032.2+4128e and 3FGL~J2032.2+4126 shows that, below 1~GeV, the extrapolated flux from FGL~J2032.2+4128e would be stronger than 3FGL~J2032.2+4126.  
It is likely, therefore, that at low energies some of the emission from 3FGL~J2032.2+4126 is being included in the measured flux from FGL~J2032.2+4128e.
A joint fit to the \veritas\ and \LAT\ data points for \TeVJ/FGL~J2032.2+4128e is well fit with a power law of index \nstat{2.39}{0.03} and normalization \fstat{3.61}{0.21}{-10} at 4.04~GeV.

It is likely that both the HE and VHE emission share a common origin. 
Given the proximity of the emission from both sources to the pulsar PSR J2032+4127, a PWN origin of this emission is a strong possibility. 
It has been suggested that PSR J2032+4127 is in a long-period binary system and that this is the origin of at least some of the VHE emission and could be confirmed through the detection of correlated variability across multiple wavelengths \cite{2032Binary}.

\subsection{\GCyg}
In this analysis, \GCyg\ was observed by \veritas\ at a peak significance of 7.6$\sigma$ using the \ext\ integration region.
The centroid was ($l$, $b$) = (\dee{78.30}{0.02}{0.01}, \dee{2.55}{0.01}{0.01}) ((\ra, \dec) = \hms{20}{20}{04.8}, \dms{40}{45}{36}) with a semi-major axis of \dee{0.29}{0.02}{0.02} and a semi-minor axis of \dee{0.19}{0.01}{0.03} at an angle of \dee{176.7}{0.1}{2} east of galactic north. 
The \LAT\ morphology of 3FGL J2021.0+4031e was examined by producing a TS map of the region with 3FGL J2021.0+4031e removed from the model of the region (\cref{fig:FermiGammaCyg}).
This shows a disk-like structure of similar size to the remnant detected in the CGPS 1420~MHz survey \cite{CGPS} but enhanced at the northern rim where the \veritas\ emission is detected.

The \veritas\ emission was fit with a power law of spectral index \nee{2.79}{0.39}{0.20} and a normalization of \fee{5.01}{0.93}{1.00}{-16} at 1500 GeV. 
Two \LAT\ SEDs were produced for 3FGL J2021.0+4031e.
The first uses the full disk of radius 0.63\degree\ as in the 3FGL catalog and has a TS of 909.44 and spectral parameters of $\alpha$ = \nstat{2.02}{0.03}, \No\ = \fstat{2.50}{0.10}{-13} at \Eo\ = 6780 GeV.
For the second 3FGL J2021.0+4031e was broken up into two sources: the region inside the \veritas\ \ext\ integration region and the remainder of the larger region outside of the \veritas\ integration region.
The two component are detected at TSs of 186.0 and 473.9 with fit parameters $\alpha$ = \nstat{2.00}{0.07}, \No\ = \fstat{5.77}{0.52}{-14} at \Eo\ = 6775 GeV, and $\alpha$ = \nstat{2.02}{0.04}, \No\ = \fstat{1.84}{0.10}{-13} at \Eo\ = 6775 GeV respectively (\cref{fig:GammaCygniSpectrum}).
A joint fit was conducted to the \veritas\ and the \LAT\ spectral points for the part that lies within the \veritas\ integration region.
The best fit is with a broken power law ($\chi^2$ = 7.8 for 9 DoF) of parameters \No\ = \fstat{1.93}{0.50}{-14}, $E_b$ = 405 GeV, $\alpha_1$ = \nstat{1.97}{0.07} and $\alpha_2$ = \nstat{2.79}{0.22}.
Since the \LAT\ emission traces the morphology of G78.2+2.1, this is the most likely origin of this emission, and thus the most likely origin of the \veritas\ emission, with the current \veritas\ observations only able to detect the brightest part of the remnant.

\subsection{\cisne}
\cisne\ was observed at a peak (local) significance  of 10.3$\sigma$ using the \ext\ integration region.
The best fit morphology ($\chi^2$/DoF = 0.838) has a centroid at ($l$, $b$) = (\dee{74.97}{0.02}{0.01}, \dee{0.35}{0.01}{0.01}) ((ra,dec) = (\hms{20}{19}{23}, \dms{36}{46}{44})) with a 1$\sigma$ angular extension of \dee{0.34}{0.02}{0.01} by \dee{0.14}{0.01}{0.02} at an angle \dee{127.0}{2.6}{0.1} east of galactic north.
The \point\ integration region analysis shows that the emission appears to be strongest in two regions, each of which is detected at a level greater than 7$\sigma$.
This suggests that the emission may be the result of two sources that were previously unresolved (though whether these are two independent sources or two enhancements of a single, extended source, is as yet, uncertain).
Jointly fitting the uncorrelated excess sky map with two, symmetric Gaussian functions identifies the sources at locations ($l$, $b$) = (\dee{74.87}{0.01}{0.01}, \dee{0.42}{0.01}{0.01}) and (\dee{75.13}{0.01}{0.01}, \dee{0.19}{0.01}{0.01}) ((ra,dec) = (\hms{20}{18}{48}, \dms{36}{44}{24}) and (\hms{20}{20}{31}, \dms{36}{49}{12})) thus we name these sources \cisneA\ and \cisneB.
The two sources are separated by 0.35\degree\ and have 1$\sigma$ extensions of \dee{0.18}{0.01}{0.04} and \dee{0.03}{0.01}{0.01} respectively.

To explore this in more detail a section through the \point\ integration region excess map along the major axis of the fit to \cisne\ was produced.
The data was binned using rectangles of width 0.05\degree\ and height 0.2\degree\ along a length of 1.0\degree.  
This section was fit with three models: a single Gaussian function, the sum of two Gaussian functions, and the sum of three Gaussian functions with p-values of 6$\times 10^{-5}$, 4$\times 10^{-8}$ and 0.12 respectively. 
The third Gaussian improved the fit beyond \cisneA\ where there is a region of excess at around the 4$\sigma$ level.

A SED for \cisne\ was produced using the \ext\ integration region and was fit with both a power law and a log parabola, with the log parabola favored but not significantly (F-test = 1.75$\sigma$).
The power law fit parameters are $\alpha$ = \nee{1.98}{0.09}{0.20} and \No\ = \fee{1.02}{0.11}{0.20}{-16} at \Eo\ = 3110 GeV with a $\chi^2$ of 8.73 for 5 DoF.
Spectra were also produced for \cisneA\ and \cisneB\ using the \point\ integration region.
\cisneA\ is fit with a power law of parameters \No\ = \fee{5.12}{0.94}{1.48}{-17} at \Eo\ = 2710 GeV and a spectral index of \nee{2.00}{0.21}{0.2}.
\cisneB\ shows a harder spectrum with a spectral index of \nee{1.71}{0.26}{0.2} and a flux normalization of \fee{3.00}{0.56}{0.60}{-17} at \Eo\ = 3270 GeV (\cref{fig:CisneSpectrum}).

\begin{wrapfigure}{r}{0pt}
\centering
\includegraphics[width=0.5\textwidth]{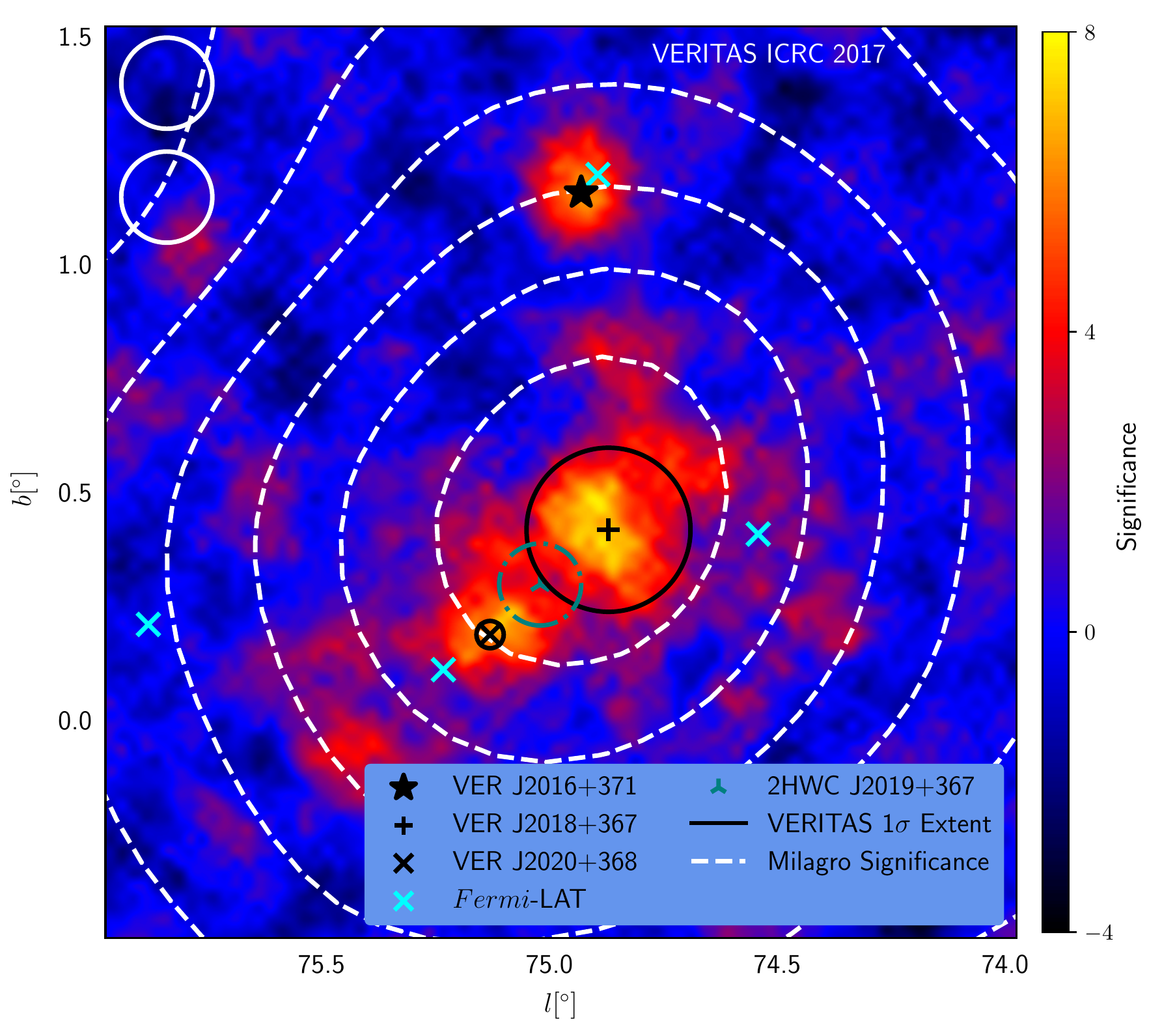}
\caption{\veritas\ significance map of the \MGROcisne\ region.  
\label{fig:VERSkymapsCisne}}
\vspace{-20pt}
\end{wrapfigure}

3FGL J2021.1\allowbreak+3651 (PSR J2021+3651) is a \GR\ pulsar that lies 0.11\degree\ away from \cisneB.
\xmm\ and \suzaku\ observations clearly show a bright point source spatially coincident with the \LAT\ detected pulsar PSR J2021.1\allowbreak+3651, with associated extended emission from a PWN (G75.2+0.1) \cite{SuzakuCisne} stretching back towards \cisneB.
It is likely that \cisneB\ is associated with G75.2+0.1.

\subsection{\CTB}
\CTB\ was observed using the \point\ integration region at 6.2$\sigma_{local}$.
The location of this fit is ($l$, $b$) = (\dee{74.94}{0.01}{0.01}, \dee{1.16}{0.01}{0.01}) ((\ra,\dec) = \hms{20}{15}{57}, \dms{37}{12}{31}) with no evidence of extension.
The spectrum is fit with a power law of normalization \fee{2.8}{1.2}{0.6}{-17} at \Eo\ = 2510 GeV and spectral index \nee{2.1}{0.8}{0.2}.

\begin{wrapfigure}{r}{0pt}
\centering
\includegraphics[width=0.5\textwidth]{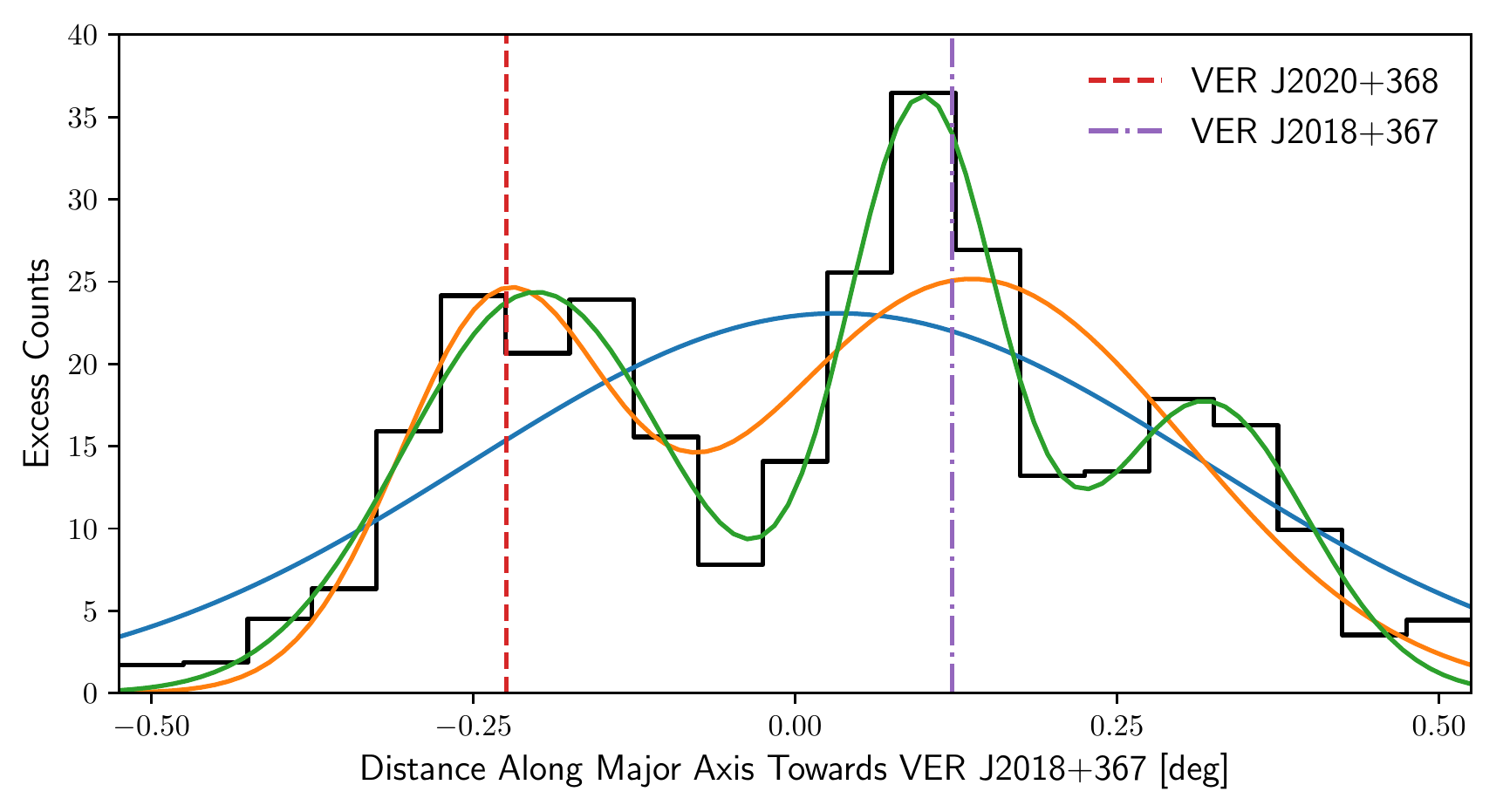}
\caption{A section along the major axis of \cisne\ towards \cisneA\ through the excess map using the \point\ integration radius.  
\label{fig:CisneTranscept}}
\end{wrapfigure}

To test the relative contributions of CTB~87 and the blazar QSO J2015+371 to the HE \GR\ emission, in addition to a single source model, a model with two power law sources at their radio locations was used.
The two sources had test statistics of 102 and 1087 respectively.
The spectra of the two sources are noticeably different, with the CTB 87 source being weaker and harder.
Plotting all three \LAT\ spectra with the \veritas\ spectrum (\cref{fig:CTB87Spectrum}), shows good agreement between the spectrum of the \LAT\ CTB 87 source and \CTB, whereas the QSO J2015+371 source would require a spectral hardening to fit the \veritas\ results.
Conducting a joint power law fit to the spectral points from the CTB 87 source and \CTB\ gives the parameters \No\ = \fstat{6.67}{0.60}{-11} at \Eo\ = 5.2 GeV and spectral index \nstat{2.39}{0.05}.
Combined, the location of \CTB\ and the spectra of the \LAT\ emission, when fit as two sources, suggests that the \fermi\ emission from the direction of CTB 87 and \CTB\ are the same source and that they are associated with the PWN CTB 87. 

\section{Comparison to \hess\ Galactic Plane Survey}
The \hess\ Galactic plane survey (GPS) \cite{HESSGPS} contains 56 sources and has similar sensitivity to this work.
In the \LAT\ 3FGL catalog there are 339 total sources in the region covered by the GPS and 37 in the region covered by this work.
We would therefore expect to detect 37/339$\times$56 = 6 VHE sources in the \veritas\ survey, comparable to the 4/5 that we detect (depending on whether \cisneA\ and \cisneB\ are considered as independent sources or as hot spots in \cisne ).
Performing similar calculations with the 2 and 3 FHL catalogs gives 3/40$\times$56 = 4 and 13/119$\times$56 = 6 VHE sources respectively.

\section{Conclusions}
The results of the \veritas\ of the Cygnus region with a total observing time of about 309 h (271 h live time) and a Pass 8 \LAT\ analysis of the same region using over 7 years (2008 August - 2016 January) of data above 1 GeV were presented.
These observations have enabled the most detailed study of the region to date in HE and VHE \GR s.
They have shown that the morphology of the HE emission from 3FGL J2021.0+4031e is not uniform, rather it peaks at the north eastern rim of the SNR, in the same region as the \veritas\ emission from \GCyg\ and the spectral fits show good agreement.
An extended (68\% containment radius = $0.15^{\circ} {}^{+0.02^\circ}_{-0.03^\circ}$) \LAT\ counterpart to \TeVJ\ (TeV J2032+4130) was detected at a test statistic of 321.
\cisne\ was resolved into two sources in the \veritas\ observations (\cisneA\ and \cisneB) with both sources are detected at a level greater than 7$\sigma_{local}$.
An exploration of the HE emission from the direction \CTB\ suggests that at least some of the emission is due to CTB~87 and a joint fit with the \veritas\ data was conducted.

\acknowledgments
We acknowledge the support listed at \url{https://veritas.sao.arizona.edu}.

\end{document}